\documentclass[5p]{elsarticle}
\pdfoutput=1

\usepackage{amssymb}
\usepackage{amsmath}
\usepackage{graphicx}
\usepackage{times}

\biboptions{sort&compress}

\journal{BBA - Proteins and Proteomics}

\begin{document}

\begin{frontmatter}

\title{How conformational changes can affect catalysis, inhibition and drug resistance of enzymes with induced-fit binding mechanism such as the HIV-1 protease}

\author[label1]{Thomas R. Weikl}

\author[label1,label2]{Bahram Hemmateenejad}

\address[label1]{Max Planck Institute of Colloids and Interfaces, Department of Theory and Bio-Systems,  Potsdam, Germany}
\address[label2]{Department of Chemistry, Shiraz University, Shiraz, Iran} 

\begin{abstract}
A central question is how the conformational changes of proteins affect their function and the inhibition of this function by drug molecules. Many enzymes change from an open to a closed conformation upon binding of substrate or inhibitor molecules. These conformational changes have been suggested to follow an induced-fit mechanism in which the molecules first bind in the open conformation in those cases where binding in the closed conformation appears to be sterically obstructed such as for the HIV-1 protease. In this article, we present a general model for the catalysis and inhibition of enzymes with induced-fit binding mechanism. We derive general expressions that specify how the overall catalytic rate of the enzymes depends on the rates for binding, for the conformational changes, and for the chemical reaction. Based on these expressions, we analyze the effect of mutations that mainly shift the conformational equilibrium on catalysis and inhibition. If the overall catalytic rate is limited by product unbinding, we find that mutations that destabilize the closed conformation relative to the open conformation increase the catalytic rate in the presence of inhibitors by a factor $\exp(\Delta\Delta G_C/RT)$ where $\Delta\Delta G_C$ is the mutation-induced shift of the free-energy difference between the conformations. This increase in the catalytic rate due to changes in the conformational equilibrium is independent of the inhibitor molecule and, thus, may help to understand how non-active-site mutations can contribute to the multi-drug-resistance that has been observed for the HIV-1 protease. A comparison to experimental data for the non-active-site mutation L90M of the HIV-1 protease indicates that the mutation slightly destabilizes the closed conformation of the enzyme.
\end{abstract}


\end{frontmatter}

\section{Introduction}

The function of proteins often involves conformational changes during the binding or unbinding of ligand molecules \cite{Gerstein98,Goh04}. Central questions are how these conformational changes are  coupled to the binding processes, and how they affect the function of the proteins and the inhibition of this function by drug molecules. For some proteins, a conformational change has been proposed to occur {\em predominantly after} a binding or unbinding process \cite{Sullivan08,Elinder10,Fieulaine11,Copeland11}, apparently `induced' by this process \cite{Koshland58}. For other proteins, a conformational change has been suggested to occur {\em predominantly prior} to a binding or unbinding process \cite{Beach05,Boehr06a,Kim07,Qiu07,Lange08,Wlodarski09,Braz10,Masterson10,Weikl12}, which has been termed `conformational selection' since the ligand appears to select a conformation for binding or unbinding \cite{Tsai99,Kumar00,Boehr09}. Binding {\em via} conformational selection implies induced-change unbinding, and {\em vice versa}, since the ordering of events is reversed in the binding and unbinding direction \cite{Weikl12}. 

\begin{figure*}[t]
\begin{center}
\hspace*{0cm}\resizebox{1.5\columnwidth}{!}{\includegraphics{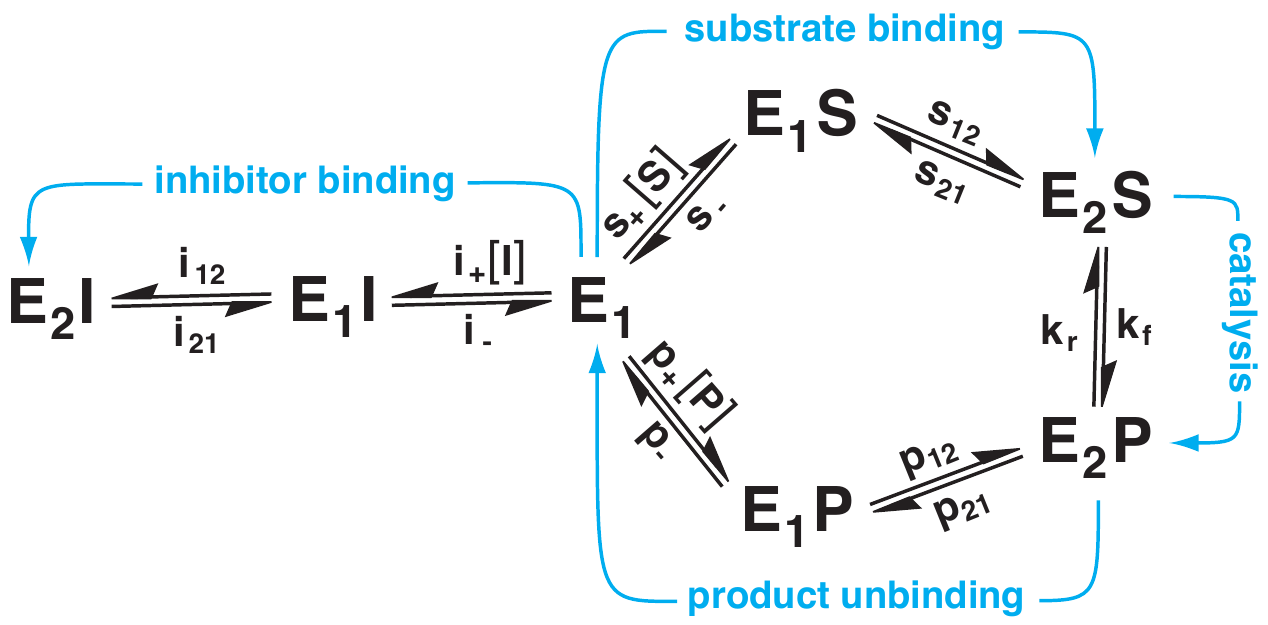}}
\end{center}
{\bf Figure 1}: 7-state model for catalysis and inhibition of an enzyme with induced-fit binding mechanism. In this model, substrate molecules $S$ and inhibitor molecules $I$ first bind in conformation $E_1$ of the enzyme. These binding events induce changes into conformation $E_2$ in which the substrate $S$ is converted into the product $P$. In the case of the HIV-1 protease, the conformation $E_1$ corresponds to the semi-open conformation, and the conformation $E_2$ to the closed conformation.
\label{figure_7states}
\end{figure*}

In this article, we extend classical models of enzyme catalysis and inhibition \cite{Cheng73} by including a conformational change during the binding and unbinding of substrate, product, or inhibitor molecules. Our aim is to investigate how the conformational change affects the catalytic rates in the presence and absence of inhibitor molecules, and how non-active-site mutations that shift the conformational equilibrium alter these catalytic rates. We focus on enzymes with induced-change binding mechanism since many enzymes close rather tightly over substrate or inhibitor molecules during binding. Binding via an induced-change mechanism, i.e.~prior to the change from the `open' to the `closed' conformation of these enzymes,  is required if the entry and exit of the ligand molecules are sterically obstructed in the closed conformation \cite{Sullivan08}. In the reverse direction, the ligands then unbind {\em via} conformational selection because the change from the closed to open conformation has to occur prior to the unbinding process. Classical models of competitive inhibition, in which the inhibitor binds to the same site as the substrate, involve four states: an empty state $E$ of the enzyme, and three states $ES$, $EP$, and $EI$ in which the enzyme is bound to a substrate molecule S, a product molecule P, or an inhibitor molecule I \cite{Cheng73}. In our extended model for enzymes with induced-change binding mechanism, the enzyme can adopt two conformations, an `open' conformation 1, and a `closed' conformation 2.  The model has seven states because induced-change binding involves an open state $E_1S$ and a closed state $E_2S$ with bound substrate, two such states $E_1P$ and $E_2P$ with bound product, and two states $E_1I$ and $E_2I$ with bound inhibitor, besides the empty state (see fig.\ 1). 

With our extended model, we derive general expressions for the catalytic rates of enzymes with induced-change binding mechanism. The catalytic rates of these enzymes depend on (i) the binding and unbinding rates of substrate, product, and inhibitor molecules in the open conformation 1, (ii) the forward and backward rates of the catalytic step, and (iii) the transition rates between the two conformations in the bound states of the enzyme (see eqs.~(\ref{MichaelisMenten}) to (\ref{pof})). Our general expressions for the catalytic rates lead to an effective four-state model with a single substrate-bound state, a single product-bound, and a single inhibitor-bound state (see fig.\ 2), but with effective on- and off-rates of substrate and product molecules that depend on the conformational transition rates (see  eqs.\ (\ref{son}) to (\ref{pof})).
  
The role of the conformational changes for catalysis and inhibition can be revealed by non-active-site mutations that slightly shift the conformational equilibrium, but do not interfere directly with binding and catalysis in the active site of the enzymes. Several groups have suggested that such shifts in the conformational equilibrium might explain why non-active-site mutations can contribute to multi-drug resistance  \cite{Maschera96,Rose98,Perryman04}, i.e.\ to  an increase of catalytic rates in the presence of different inhibitory drugs. Based on our general results for enzymes with induced-change binding mechanism, we investigate how these mutations affect catalysis and inhibition, and distinguish two cases. In case 1, the maximum catalytic rate of the enzyme is limited by the unbinding of the product. We find that the catalytic rate in the presence of inhibitors depends exponentially on the mutation-induced change $\Delta\Delta G_C$ of the free-energy difference between the two conformations of the enzyme in this case (see eqs.~(\ref{vitality_def}) and (\ref{vitality1})). Non-active-site mutations with $\Delta\Delta G_C>0$ that slightly destabilize the closed conformation 2 relative to the open conformation 1 of the enzyme lead to an increase in the catalytic rate, irrespective of the inhibitor. Such non-active-site mutations thus contribute to a multi-drug-resistance of the enzyme. In case 2, the maximum catalytic rate of the enzyme is limited by the forward rate of the catalytic step. In this case, mutation-induced changes of the conformational equilibrium have no effect on the catalytic rate in the presence of inhibitors. A comparison with experimental data for the non-active-site mutation L90M of the HIV-1 protease indicates that this enzyme appears to follow case 1, which implies that non-active-site mutations that slightly destabilize the closed conformation contribute to multi-drug resistance.

\section{Catalysis and inhibition of an enzyme with induced-fit binding mechanism}

We consider an enzyme with two conformations $E_1$ and $E_2$ that binds its substrate $S$ {\em via} an induced-change mechanism $E_1 \rightleftharpoons E_1 S \rightleftharpoons E_2 S$. We assume that the catalytic step occurs in conformation 2 of the enzyme, and that the inhibitor $I$ binds to the same site as the substrate (`competitive inhibition').  The catalytic cycle of the enzyme and the inhibition of this cycle then can be described by the 7-state model shown in fig.~1.

The catalytic rate depends on the 14 forward and backward rates between the 7 states of the model. These rates are: 

\noindent (i) The {\em binding rates} $s_{+}[S]$, $p_{+}[P]$, and $i_{+}[I]$ and {\em unbinding rates} $s_{-}$, $p_{-}$, and $i_{-}$ of substrate, product and inhibitor molecules in conformation 1 of the enzyme. Here, $[S]$, $[P]$, and $[I]$ denote the concentrations of these molecules. 

\noindent (ii) The forward and reverse {\em rates of the catalytic step} $k_f$ and $k_r$.

\noindent (iii) The {\em conformational transition rates} $s_{12}$, $s_{21}$, $p_{12}$, $p_{21}$, $i_{12}$ and $i_{21}$ between the substrate-bound states $E_1S$ and $E_2S$, product-bound states $E_1P$ and $E_2P$, and inhibitor-bound states $E_1I$ and $E_2I$. { We assume that} the bound states $E_2S$, $E_2P$, and $E_2I$ with conformation 2 of the enzyme are the ground-state conformations, while the conformations $E_1S$, $E_1P$, and $E_1I$ with conformation 1 of the enzyme are excited-state conformations. { This assumption is valid if experimental structures indicate that an enzyme adopts conformation 1 in its unbound state and conformation 2 in its bound states, since the experimental structures correspond to ground-state conformations.} The assumption implies 
\begin{equation}
s_{21} \ll s_{12} \;, \;\; p_{21} \ll p_{12} \\, \text{~~and~~} i_{21} \ll i_{12} 
\label{transition_rates}
\end{equation}
i.e.~the excitation rates $s_{21}$, $p_{21}$, and $i_{21}$ are much smaller than the corresponding ground-state relaxation rates $s_{12}$, $p_{12}$, and $i_{12}$.

\begin{figure}[t]
\begin{center}
\resizebox{\columnwidth}{!}{\includegraphics{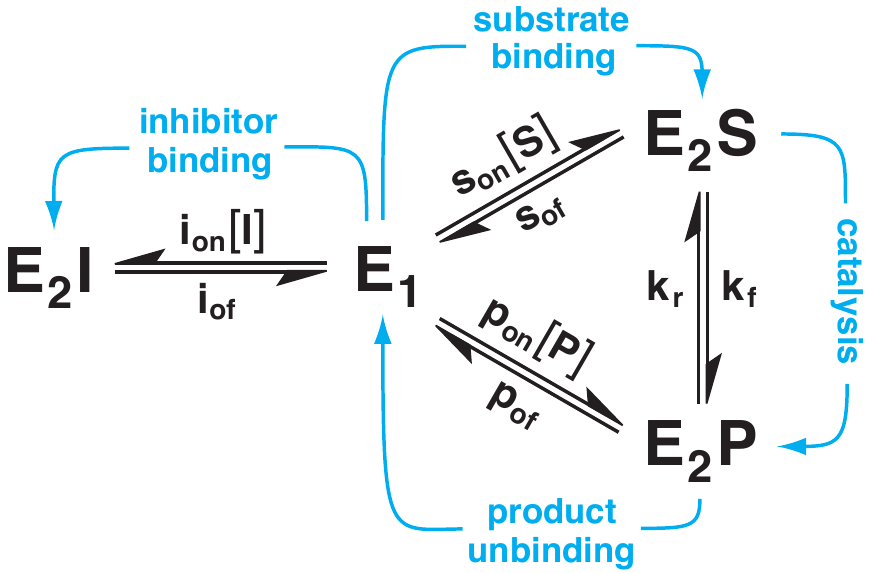}}
\end{center}
{\bf Figure 2}: Effective 4-state model for catalysis and inhibition of an enzyme with induced-fit binding mechanism. The effective binding and unbinding rates of substrate and product given in eqs.~(\ref{son}) to (\ref{pof}) connect this model to the 7-state model of fig.~1.
\label{figure_4states}
\end{figure}

One of our main results is that the catalytic rate can be written in the Michaelis-Menten form (see Appendix A) 
\begin{equation}
k \simeq \frac{k_\text{max} [S]}{(1+ [I]/K_i)K_m + [S]}
\label{MichaelisMenten}
\end{equation}
with 
\begin{equation}
k_\text{max} \simeq \frac{k_f p_\text{of}}{k_f + k_r + p_\text{of}} 
\label{kmax} 
\end{equation}
\begin{equation}
K_m \simeq \frac{(k_f + s_\text{of})p_\text{of} + k_r s_\text{of}}{(k_f + k_r + p_\text{of}) s_\text{on}} 
\label{Km} 
\end{equation}
\begin{equation}
K_i = \frac{i_{21}i_{-}}{i_\text{12}i_{+}}  
\label{Ki}
\end{equation}
and 
\begin{equation}
s_\text{on} = \frac{s_{12} s_{+}}{s_{12}+s_{-}}
\label{son} 
\end{equation} 
\begin{equation}
s_\text{of} = \frac{s_{21} s_{-}}{s_{12}+s_{-}} 
\label{sof} 
\end{equation}
\begin{equation}
p_\text{of} = \frac{p_{21} p_{-}}{p_{12}+p_{-}} 
\label{pof}
\end{equation}
for negligible product concentration $[P]$ or negligible product binding rate $p_+[P]$.  In deriving eqs.~(\ref{MichaelisMenten}) to (\ref{pof}) from the exact solution, we have assumed that the excitation rate $p_{21}$ in the product-bound state is much smaller than the relaxation rate $s_{12}$ in the substrate-bound state, besides eq.~(\ref{transition_rates}) (see Appendix A). This assumption is reasonable if the relaxation rates $s_{12}$ and $p_{12}$ in the substrate- and product-bound states are of similar magnitude, since $p_{21} \ll p_{12}$ (see eq.~(\ref{transition_rates})) then implies also $p_{21} \ll s_{12}$. 

Our general results for the 7-state model of fig.~1 lead to an effective 4-state model of catalysis and inhibition, see fig.~2. The catalytic rate of this 4-state model is described by eqs.~(\ref{MichaelisMenten}) to (\ref{Km}) with $K_i = i_\text{of}/i_\text{on}$ for negligible product concentration $[P]$. The effective 4-state model has the same structure as classical 4-state models for competitive inhibition \cite{Cheng73}, but includes two conformations for the enzyme. The rates $s_\text{on}$ and $s_\text{of}$ in eqs.~(\ref{son}) and (\ref{sof}) can be understood as the effective on- and off-rates of the substrate, i.e.~as the effective forward and backward rates between the unbound state $E_1$ and the bound ground state $E_2S$. Similarly, $p_\text{of}$ in eq.~(\ref{pof}) can be understood as the effective unbinding rate of the product from the bound ground state $E_2P$. The expressions in eqs.~(\ref{son}) and (\ref{pof}) for the effective on- and off-rates result here from an analysis of steady-state catalytic rates, but are identical with previously derived expressions for the effective relaxation rates of two-step binding processes that involve a conformational change \cite{Weikl09,Weikl12} (see appendix B).

\section{Effect of non-active site mutations on catalysis}

In general, mutations can affect all rate constants of the 7-state model shown in fig.~1, in particular if they are located in the binding site, or active site, of the enzyme. Non-active-site mutations, in contrast, may mainly affect the conformational equilibrium of the bound ground states and excited states. The conformational equilibrium constant $s_{21}/s_{12}$ of the two substrate-bound states $E_1S$ and $E_2S$ and the equilibrium constant $p_{21}/p_{12}$ of the product-bound states $E_1P$ and $E_2P$ depend on the free-energy energy differences between the states. These differences can be decomposed into the free-energy difference $\Delta G_C = G(E_2) - G(E_1)$ between the conformations and the differences in the binding free energy of substrate and product molecules in the two conformations of the enzyme. If a non-active-site mutation only affects the conformational free-energy difference but not the binding free-energy differences between the conformations, we have
\begin{equation}
\frac{s_{21}^\prime/s_{12}^\prime}{s_{21}/s_{12}} = \exp(\Delta\Delta G_C/RT)
\end{equation}
\begin{equation}
\frac{p_{21}^\prime/p_{12}^\prime}{p_{21}/p_{12}} = \exp(\Delta\Delta G_C/RT)
\end{equation}
where the prime indicates rates of the mutant. Here, $\Delta\Delta G_C = \Delta G_C^\prime - \Delta G_C$ is the mutation-induced change of the free-energy difference between the conformations. 

If the relaxation rates $s_{12}$ and $p_{12}$ into the bound ground states are much larger than the corresponding unbinding rates  $s_{-}$ and $p_{-}$ from the excited states, we obtain 
\begin{equation}
\frac{s_\text{on}^\prime}{s_\text{on}}   \simeq 1
\label{son_ratio}
\end{equation}
\begin{equation}
\frac{s_\text{of}^\prime}{s_\text{of}} \simeq \exp(\Delta\Delta G_C/RT)
\label{sof_ratio}
\end{equation}
\begin{equation}
\frac{p_\text{of}^\prime}{p_\text{of}} \simeq \exp(\Delta\Delta G_C/RT) 
\label{pof_ratio}
\end{equation}
from eqs.~(\ref{son}) to (\ref{pof}), which simplify to $s_\text{on} \simeq s_{+}$, $s_\text{of} \simeq s_{21} s_{-}/s_{12}$, and $p_\text{of} \simeq p_{21} p_{-}/p_{12}$ in this case. We have assumed here that the non-active-site mutation does not affect the binding and unbinding rate constants $s_{+}$, $s_{-}$, and $p_{-}$. Alternatively, the eqs.~(\ref{son_ratio}) and (\ref{pof_ratio}) follow from the eqs.~(\ref{son}) to (\ref{pof}) if we assume that the mutation-induced change of the conformational free-energy difference mainly affects the excitation rates, which implies $s_{21}^\prime/s_{21} \simeq \exp(\Delta\Delta G_C/RT)$ and $s_{12}^\prime \simeq s_{12}$ for the conformational transition rates of the substrate-bound states, and $p_{21}^\prime/p_{21}$ $ \simeq \exp(\Delta\Delta G_C/RT)$ and $p_{12}^\prime \simeq p_{12}$ for the transition rates of the product-bound states \cite{Weikl12}. The non-active-site mutation thus changes the effective off-rates of substrate and product molecules by the same factor$ \exp(\Delta\Delta G_C/RT)$, irrespective of the binding energies of these molecules, while the effective on-rates remain the same. 
 
With eqs.~(\ref{son_ratio}) to (\ref{pof_ratio}),  we can now determine the effect of a non-active-site mutation on the catalytic quantities $k_\text{max}$ and $K_m$ of the enzyme, which are given in eqs.~(\ref{kmax}) and (\ref{Km}). We assume that the non-active-site mutation does not affect the forward and backward rates $k_f$ and $k_r$ of the catalytic step, and consider two cases:

\subsection*{Case 1: $k_{f}$ much larger than $p_\text{of}$ and $s_\text{of}$}

If the forward rate $k_f$ of the catalytic step is much larger than the effective off-rates of product and substrate, we obtain
\begin{equation}
k_\text{max} \simeq \frac{k_f p_\text{of}}{k_f + k_r} 
\label{kmax1}
\end{equation}
\begin{equation}
K_m \simeq \frac{k_f p_\text{of} + k_r s_\text{of}}{(k_f + k_r) s_\text{on}} 
\label{Km1}
\end{equation}
from eqs.~(\ref{kmax}) and (\ref{Km}). The effect of the non-active-site mutation on $k_\text{max}$ and $K_m$ is then characterized by (see  eqs.~(\ref{son_ratio}) to (\ref{pof_ratio}))
\begin{equation}
\frac{k_\text{max}^\prime}{k_\text{max}} \simeq \frac{p_\text{of}^\prime}{p_\text{of}}\simeq \exp(\Delta \Delta G_C/RT)
\label{kmax1_ratio}
\end{equation}
and
\begin{equation}
\frac{K_m^\prime}{K_m} \simeq \frac{k_f p_\text{of}^\prime + k_r s_\text{of}^\prime}{k_f p_\text{of} + k_r s_\text{of}} \simeq  \exp(\Delta \Delta G_C/RT)
\label{Km1_ratio}
\end{equation}
for $k_f^\prime \simeq k_f$ and  $k_r^\prime \simeq k_r$. The non-active-site mutation thus changes $k_\text{max}$ and $K_m$ by the same factor $\exp(\Delta \Delta G_C/RT)$. The ratio $k_\text{max}/K_m$ of these quantities therefore is not affected by the mutation:
\begin{equation}
\frac{k_\text{max}^\prime}{K_m^\prime} \simeq \frac{k_\text{max}}{K_m} 
\label{kmaxKm1_ratio}
\end{equation}
\subsection*{Case 2: $k_f$ and  $k_r$ much smaller than $p_\text{of}$ and $s_\text{of}$}

If the rates $k_f$ and $k_r$ for the catalytic step are much smaller than the effective off-rates of substrate and product, we obtain
\begin{equation}
k_\text{max} \simeq k_f
\label{kmax2}
\end{equation}
\begin{equation}
K_m \simeq \frac{s_\text{of}}{s_\text{on}} 
\label{Km2}
\end{equation}
from eqs.~(\ref{kmax}) and (\ref{Km}). The effect of the non-active-site mutation then can be described by
\begin{equation}
\frac{k_\text{max}^\prime}{k_\text{max}} \simeq 1
\label{kmax2_ratio}
\end{equation}
and
\begin{equation}
\frac{K_m^\prime}{K_m} \simeq \frac{s_\text{of}^\prime}{s_\text{of}} \simeq 
\exp(\Delta \Delta G_C/RT)
\label{Km2_ratio}
\end{equation}
The non-active-site mutation thus changes $K_m$ by the same factor as in case 1, but does not affect the maximal catalytic rate $k_\text{max}$. From these equations, we obtain
\begin{equation}
\frac{k_\text{max}^\prime/K_m^\prime}{k_\text{max}/K_m} \simeq \exp(-\Delta \Delta G_C/RT)
\label{kmaxKm2_ratio}
\end{equation}
\section{Effect of non-active-site mutations on inhibition}

To quantify the effect of a non-active-site mutation on the interplay of catalysis and inhibition, we focus now on substrate and inhibitor concentrations $[I]$ and $[S]$ with $[I] \gg K_i $ and $[I]K_m/K_i \gg [S]$. For such concentrations, a substantial fraction of the enzyme is bound to the inhibitor, and the catalytic rate of the enzyme is (see eq.~(\ref{MichaelisMenten}))
\begin{equation}
k \simeq \frac{k_\text{max} K_i [S]}{K_m [I]}
\end{equation}
and, thus, proportional to the inhibition constant $K_i$ and inversely proportional to the inhibitor concentration $[I]$. For $[I] \ll K_i $ or $[I]K_m/K_i \ll [S]$, in contrast, the catalytic rate of the enzyme is independent of $[I]$ and $K_i$ (see eq.~(\ref{MichaelisMenten})). If the mutant also fulfills the conditions $[I] \gg K_i^\prime$ and $[I]K_m^\prime/K_i^\prime \gg [S]$ at the considered substrate and inhibitor concentrations, the ratio of the catalytic rates of mutant and wildtype is 
\begin{equation}
\frac{k^\prime}{k} \simeq \frac{K_i^\prime k_\text{max}^\prime/K_m^\prime}{K_i k_\text{max}/K_m}
\label{vitality_def}
\end{equation}
The expression on the right-hand side of this equation has been termed the `vitality' of the mutant \cite{Gulnik95}. The vitality depends on the catalytic quantities $k_\text{max}$ and $K_m$ and on the inhibition constant $K_i$ of the wildtype and mutant enzyme.

The inhibition constant $K_i$, defined in eq.~(\ref{Ki}), is the dissociation constant of the inhibitor and the enzyme. The constant depends on the free-energy difference $G(E_2I) - G(E_1)$ between the  bound ground state $E_2I$ and the unbound state $E_1$. This free-energy difference can be composed into the free-energy difference $\Delta G_C = G(E_2) - G(E_1)$ between the conformations of the enzyme, and the binding energy $\Delta G_I = G(E_2 I) - G(E_2)$ in conformation 2 of the enzyme. The effect of a mutation on $K_i$ thus can be characterized by
\begin{equation}
\frac{K_{i}^\prime}{K_i} = \exp((\Delta\Delta G_C+\Delta\Delta G_I)/RT)
\label{Ki_ratio}
\end{equation}
where $K_i^\prime$ is the inhibition constant of the mutant, and $\Delta\Delta G_C$ and $\Delta\Delta G_I$ are the mutation-induced changes of the conformational free-energy difference and the binding free energy in conformation 2. Besides affecting $\Delta G_c$, some non-active-site mutations that contribute to inhibitor resistance of an enzyme have been suggested to have an indirect effect on the binding free energy $\Delta G_I$ by slightly changing the shape of the binding pocket in a way that does not interfere with substrate binding \cite{Prabu02,Altman08,Nalam10}. This suggestion concerns inhibitors with a bound shape that is not confined within the shape or `envelope' of the bound substrate. 

With eq.~(\ref{Ki_ratio}) and the results of the previous section, we can now determine the vitality of non-active site mutants. We consider again the two cases of the previous section: 

\subsection*{Case 1: $k_{f}$ much larger than $p_\text{of}$ and $s_\text{of}$}

If the forward rate of the catalytic step is much larger than the effective off-rates of product and substrate for both wildtype and mutant, we obtain the vitality
\begin{equation}
\frac{K_i^\prime k_\text{max}^\prime/K_m^\prime}{K_i k_\text{max}/K_m} \simeq   \exp((\Delta\Delta G_C+\Delta\Delta G_I)/RT)
\label{vitality1}
\end{equation}
from the eqs.~(\ref{kmaxKm1_ratio}) and (\ref{Ki_ratio}). The effect of the non-active-site mutation on the free-energy difference between the conformation thus is reflected by the factor  $\exp(\Delta\Delta G_C/RT)$ in the vitality, while the effect of the mutation on the binding-free energy of the inhibitor in conformation 2 of the enzyme is reflected by the factor  $\exp(\Delta\Delta G_I/RT)$.  Eq.~(\ref{vitality1}) indicates that the change of the conformational free-energy difference induced by a non-active-site mutation can have a significant effect on the catalytic rate of the mutant in the presence of the inhibitors. 

\subsection*{Case 2: $k_f$ and  $k_r$ much smaller than $p_\text{of}$ and $s_\text{of}$}

If the rates for the catalytic step are much smaller than the effective off-rates of substrate and product, we obtain the vitality
\begin{equation}
\frac{K_i^\prime k_\text{max}^\prime/K_m^\prime}{K_i k_\text{max}/K_m}  \simeq  \exp(\Delta\Delta G_I/RT)
\label{vitality2}
\end{equation}
from the eqs.~(\ref{kmaxKm2_ratio}) and (\ref{Ki_ratio}). In this case, the vitality is only affected by the change $\Delta\Delta G_I$ in the binding-free energy of the inhibitor.

\section{Comparison with experimental data for the non-active site mutation L90M of the HIV-1 protease}

To illustrate our general results for the effect of non-active-site mutations on catalysis and inhibition of enzymes with induced-fit binding kinetics, we consider here the well-studied mutation L90M of the HIV-1 protease as an example.
The HIV-1 protease changes from a semi-open to a closed conformation during binding of substrate or inhibitor molecules. The conformational change from the semi-open to the closed conformation has been suggested to occur after the binding process since binding in the closed conformation appears to be sterically prohibited \cite{Furfine92,Gustchina90}. The conformational dynamics of the HIV-1 protease has been studied in NMR experiments \cite{Nicholson95,Ishima99,Freedberg02}, by pulsed EPR spectroscopy \cite{Galiano07,Torbeev09,Blackburn09},  and in molecular dynamics simulations \cite{Collins95,Scott00,Piana02,Perryman04,Hornak06,Wiley09,Sadiq10,Genoni10,Deng11}. 
In the native state, the HIV-1 protease is a homo-dimer of two identical subunits with 99 residues. The active site is located in a tunnel at the dimer interface, which opens and closes {\em via} the motion of two `flaps', accompanied by motion in other regions of the protease.
Inhibitors of the HIV-1 protease that bind to the active site play a central role in anti-AIDS therapies. However, the efficiency of these therapies is impaired by viral mutations that lead to inhibitor resistance. Mutations that are associated with resistance occur both in the active site of the HIV-1 protease and distal to the active site. While active-site mutations can directly interfere with catalysis and binding, the contribution of distal, non-active-site mutations to resistance is not fully clear. Several  groups have suggested that non-active-site mutations of the HIV-1 protease such as L90M contribute to inhibitor resistance by shifting the equilibrium between the semi-open and closed conformation of the protease \cite{Maschera96,Rose98,Perryman04}. Alternative explanations for the mutation L90M include an indirect effect on active-site residues \cite{Hong00,Mahalingam01,Ode06,Kovalevsky06}, and a reduction of dimer stability \cite{Xie99} since the mutated residue 90 is located at the dimer interface.  

\begin{table}
{\normalsize {\bf Table 1}: Catalytic efficiency of mutant L90M relative to wildtype HIV-1 protease (data from Maschera et al.\cite{Maschera96})}
\label{table_catalysis}
\begin{center}
\begin{tabular}{lccc}
\hline\\[-0.3cm]
substrate & $\displaystyle \frac{k_\text{max}^\prime}{k_\text{max}}$ & $\displaystyle \frac{K_m^\prime}{K_m}$ &  $\displaystyle \frac{k_\text{max}^\prime/K_m^\prime}{k_\text{max} /K_m}$  \\[0.3cm]
\hline \\[-0.3cm]
SQNY-PIVQ & 1.3 & 1.4 & 0.9\\
ARVL-AEAM & 1.6 & 1.3 & 1.3 \\
ATIM-MQRG & 1.4 & 1.6 & 0.9 \\
PGNF-LQSR & -- & -- & 0.4 \\
SFNF-PQIT & 1.3 & 1.9 & 0.8 \\
TLNF-PISP & 1.2 & 1.1 & 1.1 \\
AETF-YVDG & 1.4 & 1.2 & 1.2 \\
RKIL-FLDG &  1.2 &  0.9 & 1.5 \\
fluorogenic & 1.4 & 2 & 0.7 \\[0.1cm]
\hline\\[-0.3cm]
\end{tabular}
\end{center}
\end{table}

Maschera and co-workers \cite{Maschera96} have systematically investigated the catalytic efficiency  of the wildtype protease and the mutant L90M with 8 different peptide substrates and a fluorogenic substrate (see Table 1). The mutation L90M leads to a slight increase of $k_\text{cat}$ and $K_m$. Averaged over all substrates, the ratios of the experimental values for the mutant and the wildtype are $k_\text{max}^\prime/k_\text{max}=1.35 \pm 0.05$ and $K_m^\prime/K_m = 1.43\pm 0.14$, while the average value for the ratio of $k_\text{max}^\prime/K_m^\prime$ and $k_\text{max}/K_m$ is $0.98 \pm 0.12$. These results are consistent with eqs.~(\ref{kmax1_ratio}) to (\ref{kmaxKm1_ratio}), which indicate a change of $k_\text{cat}$ and $K_m$ by the same factor  $\exp(\Delta \Delta G_C/RT)$ due to a non-active-site mutation that mainly affects the conformational equilibrium.  From the average values of $k_\text{max}^\prime/k_\text{max}$ and  $K_m^\prime/K_m$ obtained from the data of Maschera et al., the change in the conformational free-energy difference $\Delta \Delta G_C$ due to the mutation L90M can be estimated as about 0.2 kcal/mol, which indicates a slight destabilization of the closed conformation of the HIV-1 protease relative to the semi-open conformation. However, other groups have reported a decrease of $k_\text{max}$, $K_m$, and $k_\text{max}/K_m$ due to the mutation L90M for three different substrates \cite{Mahalingam04,Kozisek07,Nillroth97}. In general, the catalytic efficiency of the HIV-1 protease strongly depends on the experimental conditions, in particular on the pH and salt concentration \cite{Szeltner96,Maschera96,Muzammil03}. 

\begin{table}[t]
{ {\bf Table 2}: Ratio of dissociation constants $K_D^\prime$ and $K_D$ for the mutant L90M and the wildtype of the HIV-1 protease (calculated from data of Shuman et al.\cite{Shuman03})}
\label{table_KD}
\begin{center}
\begin{tabular}{lcc}
\hline\\[-0.4cm]
inhibitor &~~~~~~~&  $K_D^\prime / K_D$  \\[0.1cm]
\hline
Amprenavir  && $2.73 \pm 0.67$\\
Indinavir && $2.85\pm 0.71$\\
Nelfinavir && $1.99 \pm 0.70$ \\
Ritonavir && $4.54\pm 1.68$ \\
Saquinavir  && $2.93\pm 0.77$ \\
\hline
\end{tabular}
\end{center}
\end{table}

Shuman et al.~\cite{Shuman03} have measured the dissociation constants $K_D$ and $K_D^\prime$ of the wildtype and the mutant L90M for five different inhibitors with biosensor methods (see Table 2). Within the experimental errors, the ratios $K_D^\prime/K_D$ of the dissociation constants for the five inhibitors agree with the average value $3.0 \pm 0.4$ of these ratios. For a non-active-site mutation with a negligible change $\Delta\Delta G_I$ of the binding free energy in the closed conformation of the protease, we expect a change of $K_D$ by the inhibitor-independent factor $\exp(\Delta\Delta G_C/RT)$ from eq.\ (\ref{Ki_ratio}) since $K_D$ is identical with the inhibition constant $K_i$. From the mean value $K_D^\prime/K_D\simeq 3$, we obtain the estimate $\Delta\Delta G_C\simeq 0.65$ kcal/mol for $\Delta\Delta G_I \simeq 0$, which is somewhat larger than the estimate $\Delta\Delta G_C\simeq 0.2$ kcal/mol obtained from the changes of $k_\text{max}$ and $K_m$ reported by Maschera et al. (see above). 

Inhibition constants $K_i$ and $K_i^\prime$ for the wildtype and mutant L90M have been measured by several groups for a variety of inhibitors (see Table 3). The ratios $K_i^\prime/K_i$ differ for different inhibitors, which appears to indicate mutation-induced changes $\Delta\Delta G_I$ in the binding free energy for the closed conformation that depend on the inhibitor. However, results by different groups for the same inhibitor also differ (see data in Table 3 for the inhibitors Indinavir and Saquinavir), possibly due to different experimental conditions such as pH or salt concentrations. Overall, the majority of the experimentally determined ratios $K_i^\prime/K_i$ for L90M in Table 3 are larger than 1, which may be interpreted to point towards a destabilization of the closed conformation relative to the semi-open, i.e.~towards positive values of $\Delta\Delta G_C$.

\begin{table}[t]
{\bf Table 3}: Ratio of inhibition constants $K_i^\prime$ and $K_i$ for the mutant L90M and the wildtype of the HIV-1 protease
\label{table_Ki}
\begin{center}
\begin{tabular}{l c  c}
\hline
inhibitor & reference &  $K_i^\prime / K_i$  \\[0.1cm]
\hline
Indinavir & \cite{Kozisek07} & $4.1 \pm 0.8$ \\
  & \cite{Kovalevsky06} & $1.3 \pm 0.3$ \\
  & \cite{Mahalingam04}  & $0.16 \pm 0.03$ \\
Nelfinavir & \cite{Kozisek07}  & $9.0 \pm 1.5$ \\
Ritonavir & \cite{Kozisek07}  & $3.2 \pm 1.0$\\
Saquinavir & \cite{Maschera96} & $21\pm 4$ \\
       &  \cite{Kozisek07} & $28 \pm 9$ \\
       & \cite{Ermolieff97} & $3.0\pm 1.0$ \\
L-735,524 & \cite{Maschera96} & $5.8 \pm 1.7$ \\
VX-478 & \cite{Maschera96} & $2.7 \pm 0.3$\\
AG-1343 & \cite{Maschera96} & $3.5 \pm 1.0$ \\
 ABT-538 & \cite{Maschera96}  & $6.7 \pm 2.4$ \\
 Lopinavir &  \cite{Kozisek07}  & $1.0 \pm 0.6$ \\
QF34 &  \cite{Kozisek07} & $0.9 \pm 0.4$\\
Darunavir (TMC-114) & \cite{Kovalevsky06} &  $0.14 \pm 0.04$ \\
\hline
\end{tabular}
\end{center}
\end{table}

\section{Conclusions}

We have considered here a general 7-state model for the catalysis and inhibition of enzymes with induced-fit binding mechanism. This model extends classical models for catalysis and inhibition \cite{Cheng73} by including changes between two different conformations of the enzymes, and helps to understand how non-active-site mutations that shift the conformational equilibrium can affect catalysis and inhibition. The induced-fit binding mechanism considered here applies to enzymes such as the HIV-1 protease that close over ligands in the bound ground-state conformation in a way that sterically prevents ligand exit and entry in this conformation \cite{Sullivan08,Furfine92,Gustchina90}. The unbinding of product and inhibitor molecules from these enzymes follows a conformational-selection mechanism since a conformational change from the closed ground-state conformation to an open or semi-open exited-state conformation is required prior to ligand exit \cite{Weikl12}.  Our approach can be generalized to enzymes with other binding mechanisms, e.g.~to enzymes with a conformation-selection binding and induced-fit unbinding mechanism.

\section*{Acknowledgements}

Support from the Alexander von Humboldt Foundation {\em via} a Georg Forster Research Fellowship for Bahram Hemmateenejad and from the Shiraz University Research Council are gratefully acknowledged.

\appendix
\section{Catalytic rates}

We illustrate in this section how eqs.~(\ref{MichaelisMenten}) to (\ref{pof}) for the catalytic rate of the 7-state model shown in fig.~1 can be derived from exact results. To avoid exceedingly long equations, we focus on the special case with inhibitor concentration $[I] = 0$. The derivation in the general case with $[I] > 0$ is analogous to the derivation in this special case.  

For $[I] = 0$, the two states $E_1I$ and $E_2I$ have zero probability. The 7-state model thus reduces to a model with 5 states. The steady-state probabilities $P(E_1)$, $P(E_1S)$, $P(E_2S)$, $P(E_2P)$, and $P(E_1P)$ of these 5 states can be calculated from the 5 equations (see e.g.~\cite{Hill89})
\begin{equation}
P(E_1)+P(E_1S)+P(E_2S)+P(E_2P)+P(E_1P)=1
\label{eq1}
\end{equation}
\begin{equation}
P(E_1S) s_{-} + P(E_1P) p_{-} - P(E_1) s_{+}[S] = 0
\label{eq2}
\end{equation}
\begin{equation}
 P(E_1) s_{+}[S]+ P(E_2S) s_{21} - P(E_1S) (s_{12}+ s_{-}) = 0
\end{equation}
\begin{equation}
 P(E_1S) s_{12}+ P(E_2P) k_r - P(E_2S) (s_{21}+ k_f) = 0
\end{equation}
\begin{equation}
P(E_2S)k_f + P(E_1P) p_{12} - P(E_2P)(k_r + p_{21})=0
\label{eq5}
\end{equation}
While eq.~(\ref{eq1}) ensures probability normalization, the eqs.\ (\ref{eq2}) to (\ref{eq5}) are the flux balance conditions for the states $E_1$, $E_1S$, $E_2S$, and $E_2P$ in the steady state, in which the inward flux into a state equals the outward flux from this state. We have assumed that the product concentration $[P]$ or product binding rate $p_+[P]$ are negligible in eq.~(\ref{eq2}).

The catalytic flux $k$ of the enzyme is defined as the steady-state flux along the cycle:
\begin{equation}
k = P(E_2S) k_f - P(E_2P) k_r
\label{catalytic_flux}
\end{equation}
From eqs.~(\ref{eq1}) to (\ref{catalytic_flux}), we obtain the exact result for catalytic flux
\begin{equation}
k = \frac{a [S]}{b + c [S]}
\end{equation}
with
\begin{eqnarray}
a  &=& s_{+} s_{12} k_f p_{21} p_{-}
\label{a_exact}
\\
b &=& k_f p_{21} p_{-} (s_{12}+s_{-}) + k_r s_{21} s_{-}(p_{12}+p_{-}) \nonumber\\
&&+\, s_{21}s_{-}p_{21} p_{-} 
\label{b_exact}
\\
c &=& k_f (p_{21} p_{-} + (p_{12} + p_{21} + p_{-}) s_{12})s_{+} \nonumber\\
&&+\, (k_r (p_{12}+p_{-})+ p_{21} p_{-})(s_{12}+s_{21})s_{+} 
\label{c_exact}
\end{eqnarray}
For $p_{21} \ll p_{12}$ and $s_{21} \ll s_{12}$ (see eq.~(\ref{transition_rates})), eq.~(\ref{c_exact}) can be simplified to
\begin{eqnarray}
c &\simeq& k_f ((p_{21} +s_{12})p_{-} + p_{12}   s_{12})s_{+} \nonumber\\
&&+\, (k_r (p_{12}+p_{-})+ p_{21} p_{-})s_{12}s_{+}
\end{eqnarray}
If we further assume $p_{21} \ll s_{12}$ (see text after eq.~(\ref{pof}) for justification), we obtain 
\begin{equation}
c\simeq( (k_f + k_r)(p_{12} +p_{-}) + p_{21} p_{-})s_{12}s_{+}
\label{c_simplified}
\end{equation}
The eqs.~(\ref{a_exact}) and (\ref{c_simplified}) now lead to the maximal catalytic flux
\begin{equation}
k_\text{max} = \frac{a}{c} \simeq \frac{k_f p_{21} p_{-}}{(k_f + k_r)(p_{12} +p_{-}) + p_{21} p_{-}}
\end{equation}
This expression for the maximal catalytic flux is identical to eq.~(\ref{kmax}) with $p_\text{of}$ given in eq.~(\ref{pof}). Similarly, the eqs.~(\ref{b_exact}) and (\ref{c_simplified}) lead to an expression for  $K_m = b/c$ that is identical to eq.~(\ref{Km}) with eqs.~(\ref{son}) to (\ref{pof}).

\section{Effective rates for induced-fit binding} 

We consider here the induced-fit binding process of an inhibitor $I$ to an enzyme that can adopt two conformations $E_1$ and $E_2$:
\begin{equation}
E_1 
\;
\begin{matrix} 
\text{\footnotesize $i_{+}[I] $}\\
\text{\Large $\rightleftharpoons$}\\
\text{\footnotesize $i_{-}$}
\end{matrix}
\;
E_1 I
\;
\begin{matrix} 
\text{\footnotesize $i_{12}$}\\
\text{\Large $\rightleftharpoons$}\\
\text{\footnotesize $i_{21}$} 
\end{matrix}
\;
E_2 I
\label{inhibition}
\end{equation}
In the forward direction of this process, the conformational change occurs after the binding event, apparently `induced' by this event. In the reverse direction, the conformational change precedes the unbinding step. The unbinding of the inhibitor thus follows a conformational-selection mechanism. Here,  $i_{+}$ and $i_{-}$ are the rate constants for the binding and unbinding of the inhibitor in conformation 1 of the enzyme, and $i_{12}$ and $i_{21}$ are rates for the conformational exchange between the bound ground state $E_2 I$ and the bound excited state $E_1 I$. We assume that the concentration $[I]$ of the inhibitor is much larger than the enzyme concentration, which implies pseudo-first-order kinetics with approximately constant inhibitor concentration and, thus, constant binding rate $i_{+}[I]$.

The effective on- and off-rates of the two-step process (\ref{inhibition}) can be determined from the dominant relaxation rate of this process \cite{Weikl09,Weikl12}. Since the excitation rate $i_{21}$ of the bound excited state $E_1 I$ is much smaller than the relaxation rate $i_{12}$ into the bound ground state $E_2 I$, the effective off-rate from $E_2I$ to $E_1$ is \cite{Weikl09,Weikl12}:
\begin{equation}
i_\text{of} \simeq \frac{i_{21} i_{-}}{i_{12}+i_{-}}  
\label{dof_full}
\end{equation}
The effective on-rate $i_\text{on}$ follows from the condition that the effective equilibrium constant $i_\text{on}/i_\text{of}$ of the two-step process has to be equal to the product $(i_{+}/i_{-})(i_{12}/i_{21})$ of the equilibrium constants for the two substeps. With eq.~(\ref{dof_full}), this condition leads to
\begin{equation}
i_\text{on} \simeq \frac{i_{12} i_{+}}{i_{12}+i_{-}}
\label{don_full}
\end{equation}
%





\end{document}